# ATEN-ALINDA ORBITAL EVOLUTION: THE EXAMPLE OF POSSIBLE INTERACTING ORBITS

A. Rosaev

*FGUP NPC "NEDRA"*

Minor planets 887 Alinda and 2062 Aten are typical near Earth asteroids (NEA). Our previous researches show their possible origin in catastrophic collision near Earth orbit. It is determine the target of present research – to investigate interactions between these objects in more details.

Orbits m.p. 887 and 2062 are crossed near pericenter and apocenter accordingly. So possible interval of the interaction of these objects is repeated through the significant time interval and has relatively small duration.

On the strength of several reasons, orbits of fragments of the disastrous collision can have no an exact intersection. Besides, when integrating on the greater gap in the past increases a role of the uncertainty of initial values of elements of orbits. These factors determine a probabilistic nature of the result as well as need of modeling of the disastrous process. However, on the strength of the specific spatial location of orbits, result has highly determined numeric expression - in the suggestion on the possible interaction m. p. 887 and 2062, this event has occurred about 1400 years ago.

The possible participation some other near Earth minor objects, include Quadrantides meteoric stream, in described disastrous event is discussed.

1. INTRODUCTION

Minor planets 887 Alinda and 2062 Aten are typical near earth objects ((NEA). They represent different dynamical groups of NEA, Amor's and Aten's groups accordingly. The origin of Aten's group has a particular interest due to specific character of their orbits. The elements of orbits m.p. 887 and 2062 are given in the table 1.

Orbits m.p. 887 and 2062 are crossed near pericenter and apocenter accordingly (Fig.1). So possible interval of the interaction of these objects is repeated through the significant time interval and has relatively small duration.

Table1
Aten – Alinda orbital elements

| Element | Aten | Alinda |
|---------|--------|--------|
| q | 0.79 | 1.08 |
| e | 0.18 | 0.56 |
| i | 18.93 | 9.31 |
| Ω | 107.96 | 110.04 |
| ω | 147.92 | 350.01 |

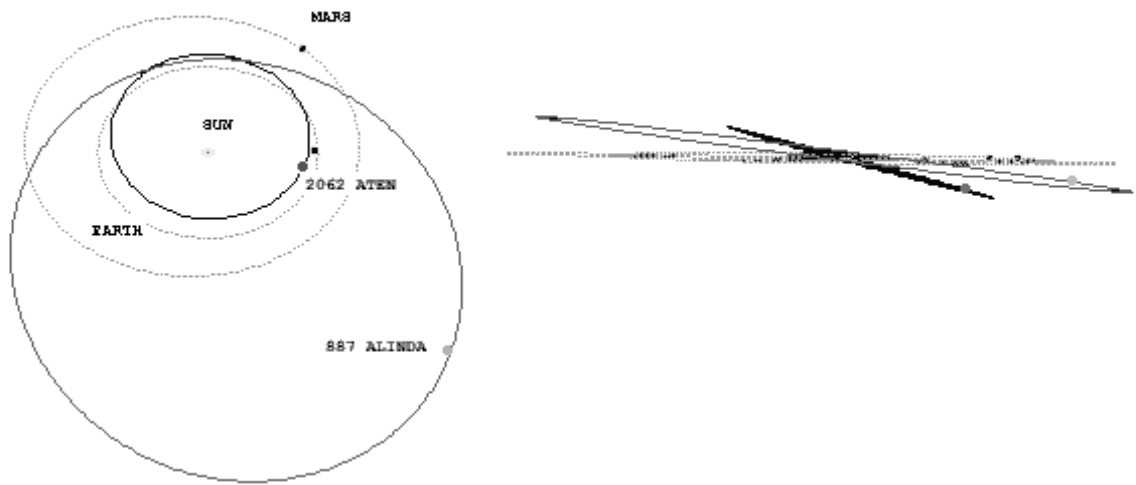

Fig.1  Aten – Alinda orbits orientation

The results of numeric integration 887 Alinda and 2062 Aten [1] from 2250 to - 4500 year, with all planetary perturbation, are used. In addition, integration at interval 1950-2010 with initial elements variation is developed. It give the ability to estimate the stability of the obtained orbits. The important role of minimal distance between NEA orbits for studying their genetic relation noted in [2].

## 2. EVOLUTION OF ATEN AND ALINDA ORBITS AND THEIR INTERACTION

The evolution of the m.p. 887 and m.p. 2062 angular orbital elements (node distance, perihelion argument and perihelion longitude, with all planetary perturbations) is constructed. It seems, that encounters with planets cannot change the character of Aten's orbit evolution during studied interval (-4500-2200 year); the character of Alinda's orbit evolution has changed about 500 year and more significantly, about -2500 year. It means, that encounters with planets cannot prevent Alinda - Aten interaction, if it really take place about 600 year. Then, variation of initial values with 70-year integration was applied. In results, the estimation of initial conditions uncertainties is obtained. It is shown, that for significant initial conditions variations, the character of orbital evolution is stable. For example, at tremendous uncertainties in incline ($0.9^o$), the deviation in n is about $0.01^o$ /50 year, or about 0.3 degree/1500 year. For real orbital elements accurance, this estimation is improved up to $0.1*10^{-5}$ degree/1500 year, or about $3*10^{-7}$ a.u. in the most poor case. It means, that uncertainties of initial conditions cannot change character Alinda - Aten interaction, if it really take place about 600 year.

The orbital distance and longitude of Aten - Alinda intersection point evolution are given at Fig. 2-3. It seems, first of all, that small time interval, exact when orbital intersection between Aten and Alinda was possible about 600 year took place. The character of orbital evolution and geometry show, that similar orbital configuration can repeat after long time interval (about 100 thousand years). It is very high possibility, that strong planetary perturbation significantly change (maybe destroy completely) circumstances of Aten - Alinda orbital intersection through this time interval. So, if interaction between m.p. 887 and m.p. 2062 really take place, it most possible time about 1400 years ago.

Velocities of the pericenter rotation, determined on results numeric integrating from accounts of all perturbations from planets on the interval 9000 years and on more short interval provided in the table 2. In spite the value of pericenter rotation strongly varying from year to year, an average tendency of secular variation maybe detected.

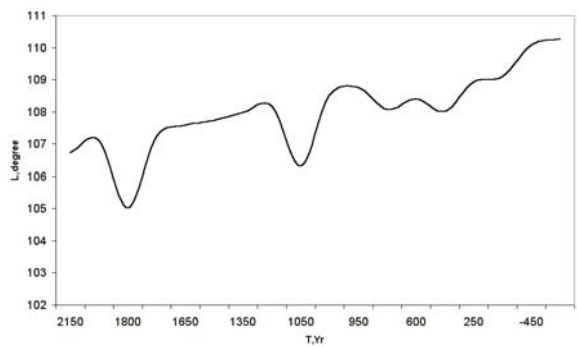

Fig. 2   Longitude of intersection point evolution

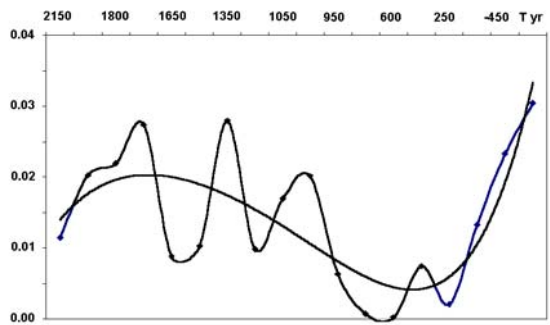

Fig. 3   Aten - Alinda orbital distance evolution

Then, the main question is how close encounters between Aten and Alinda really take place during epoch of orbital intersections. Unfortunately, the accurance of numerical integration [1] not sufficient for this purpose, so zero distance encounter between Aten and Alinda not found. There is more accurate integration (or theory of motion) is required. It need to take into account non-gravity effects, and maybe attraction of another minor planets (in Alinda case). However, there are few reasons, due to an exact intersection never observed.

Table 2
Aten – Alinda orbital elements evolution

| Epoch | ω | | Ω | |
|---|---|---|---|---|
| | Aten | Alinda | Aten | Alinda |
| 2000 | 147.91 | 350.01 | 107.96 | 110.04 |
| 1500 | 146.11 | 344.76 | 110.46 | 115.31 |
| 1000 | 144.32 | 340.84 | 112.97 | 120.13 |
| 500 | 142.39 | 334.93 | 115.49 | 126.36 |
| 0 | 140.82 | 330.83 | 118.00 | 132.32 |
| -500 | 139.19 | 325.60 | 120.53 | 138.80 |
| -1000 | 137.47 | 315.85 | 123.11 | 148.50 |
| -1500 | 135.49 | 310.13 | 125.64 | 156.40 |
| -2000 | 134.21 | 302.40 | 128.16 | 165.91 |

## 3. A FEW CIRCUMSTANCES OF POSSIBLE INTERACTIONS AND POSSIBLE RELATED OBJECTS

There are few possible reasons, lead to the observed orbital intersection:

1. Collision with comet or with another Near Earth Asteroid
2. Tidal splitting of the quickly rotating body
3. Destruction of binary asteroid by solar or Earth's perturbation

It is evident, that at described case of orbital orientation, collision breakup has small possibility. Each of the subsequent mechanisms has some addition physical phenomena, making the destruction more easy.

For the tidal splitting it is irregular form (rotation around small axis) and thermal stresses into parent body. Comets, during their evolution, become irregular shape and long time rotate around small axis.

For the destruction of binary asteroid, exact intersection of orbits not necessary. In according with recent radar researches [3], significant part of NEA may be a double systems.

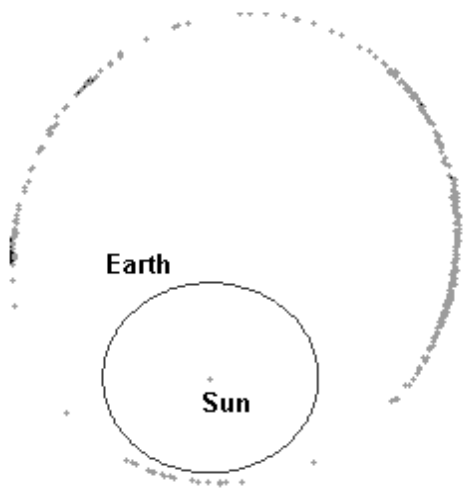

Fig. 4. Alinda-asteroid orbital intersections (along Alinda orbit)

On the strength of several reasons, orbits of fragments of the disastrous collision can have no an exact intersection. The first group of factors acts directly at the moment of destruction:

1. gravitation interaction of the fragments just after splitting
2. unknown fragments of splitting
3. rotation of parent body
4. uncertainty of observed orbital elements

The second group of factors acts more long time (during evolution):

1. differential planetary perturbation
2. subsequent destructions
3. close encounters with large planets
4. non-gravitation effects

Besides, when integrating on the greater time interval in the past a role of the uncertainty of initial values of elements of orbits increases.

These factors determine a probabilistic nature of the result as well as need of modeling of the disastrous process.

The distribution of intersection points Aten ant Alinda orbits with another minor planet (Fig.4-5) allow to suppose different collisional history and different origin of each objects. Alinda has origin in main belt, in contrary, Aten has most number of collision near Earth

However, in both cases, there is the same concentration of orbits intersection point close to longitude $l=110°$.

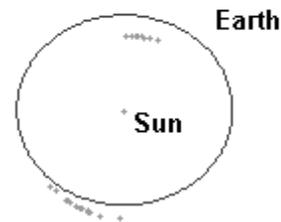

Fig. 5. Aten-asteroid orbital intersections (along Aten orbit)

Now it is impossible to exclude a participation some other near Earth minor planets in described disastrous event (table3).

The parent body for a Quadrantides meteoric stream is unknown for now. On the other hand, possible stream, related with Alinda is calculated in [4]. The elements of orbits of these streams are shown in the table 4.

Table 3
Distances between Aten orbit and another minor planet orbits

| OBJECTS | D | L | R | Z |
|---|---|---|---|---|
| Minos | 0.0035 | 117.97 | 1.081 | 0.060 |
| 1991 BA | 0.0110 | 106.30 | 1.105 | -0.000 |
| 1993 HD | 0.0239 | 107.65 | 1.103 | 0.006 |
| 1993 RA | 0.0031 | 91.66 | 1.132 | -0.099 |
| 1993 VD | 0.0053 | 113.94 | 1.089 | 0.042 |
| 1997 AC11 | 0.0187 | 127.37 | 1.049 | 0.122 |
| 1997 WT22 | 0.0051 | 130.78 | 1.045 | 0.129 |
| 1998 CS1 | 0.0040 | 107.88 | 1.106 | -0.003 |
| 1998 HL3 | 0.0019 | 101.62 | 1.117 | -0.037 |
| 1998 ST4 | 0.0052 | 92.27 | 1.131 | -0.092 |
| 1999 AF4 | 0.0091 | 110.91 | 1.097 | 0.022 |
| 1999 UR | 0.0068 | 119.85 | 1.077 | 0.067 |
| 1999 YD | 0.0078 | 111.88 | 1.093 | 0.032 |
| 2000 AC6 | 0.0033 | 111.10 | 1.098 | 0.020 |
| 2000 BH19 | 0.0130 | 111.74 | 1.095 | 0.027 |
| 2000 BK19 | 0.0091 | 118.82 | 1.078 | 0.065 |

It is possible that Quadrantides meteoric stream was formed at described catastrophic event (1400 years back). It may be related with one of the member of the group minor planets in table 3. The velocity can varying in accordance with parent body, but longitude (epoch of the stream) is fixed.

Table 4
Quadrantides and Alinda related stream orbital elements

|  | Quadrantides | Alinda related stream | Aten related stream |
|---|---|---|---|
| Epoch | 03 Jan | 04 Jan | 11 Jan |
| R.A. | 230 | 26 | 142 |
| delta | +49 | -42 | -44 |
| V | 41 | 7 | 10 |

CONCLUSIONS

The investigation of possible interaction between Aten and Alinda orbits leads to conclusion of their possible origin at breakup event close to 1400 years ago. Another similar configuration will repeat very far in past – about few hundreds thousand years of more. The catastrophic event took place close to pericenter of a parent body orbit. Now it is impossible to exclude a origin some other near Earth minor planets and Quadrantides meteor stream in described disastrous event

In all cases, problem of source of Near Earth objects appear. Some of them, like Aten, can have origin directly in Earth's neighborhood. In this relation, the construction of precision theory of motion for such objects required.